# Voltage Control of Perpendicular Exchange Bias in Multiferroic Heterostructures

*Qu Yang[1], Zhongqiang Hu[1,*], Yao Zhang[1], Wei Su[1], Bin Peng[1], Jingen Wu[1], Ziyao Zhou[1,*], Yun He[2], Wanzhao Cui[2], Zhiguang Wang[1], Ming Liu[1,*]*

Qu Yang, Prof. Zhongqiang Hu*, Yao Zhang, Wei Su, Bin Peng, Ziyao Zhou*, Zhiguang Wang, Prof. Ming Liu*
Electronic Materials Research Laboratory, Key Laboratory of the Ministry of Education, School of Electronic and Information Engineering, State Key Laboratory for Mechanical Behavior of Materials, Xi'an Jiaotong University, Xi'an 710049, China.

Yun He, Wanzhao Cui
National Key Laboratory of Science and Technology on Space Microwave, China Academy of Space Technology (Xi'an), Xi'an 710100, China

*E-mail: zhongqianghu@xjtu.edu.cn, ziyaozhou@xjtu.edu.cn, mingliu@xjtu.edu.cn;

Abstract: Perpendicular exchange bias (EB), which combines the perpendicular magnetic anisotropy and the ferromagnetic (FM) - antiferromagnetic (AFM) exchange coupling, is extremely important in the high-density AFM spintronics. However, the effective modulation of EB remains challenging, since the alternant spins at the AFM/FM interface are strongly pinned by the AFM layer. Voltage tuning of EB through the magnetoelectric coupling provides a potential way to achieve a rapid magnetization switching in an energy-efficient manner. Nevertheless, the interfacial strain mediation of perpendicular EB induced by E-field remains unexplored. In this work, we obtain perpendicular EB nanostructure by room-temperature fabrication process, and demonstrate the voltage tunable perpendicular EB in Pt/IrMn/(Co/Pt)$_2$/Ta/(011) Pb(Mg$_{1/3}$Nb$_{2/3}$)O$_3$-PbTiO$_3$ multiferroic heterostructure. To enhance the voltage control effect on perpendicular EB, we further investigate both strain-mediated magnetoelectric coupling and ionic liquid gating method in the thinned EB system with the structure of Pt/IrMn/Co/Pt/Ta. As a result, the voltage induced lattice distortion effectively transmits to the AFM/FM interface, while the charge accumulation in gating method generates a relatively large hysteresis loop offset that has not been observed before at room temperature. The voltage manipulation of perpendicular EB at room temperature provides new possibilities towards novel AFM devices and memories with great energy-efficiency and ultra-high density.

Exchange bias (EB), generated by the interlayer coupling of ferromagnetic (FM) and antiferromagnetic (AFM) materials, is one of the most decisive factors in the excellent performance of spintronic devices.[1, 2] Researches have intensely studied EB properties due to the importance applications in magnetic random access memories and magnetoresistive read heads based on spin valves or tunnel junctions.[3] If EB is combined with the perpendicular magnetic anisotropy (PMA),[1] it is promising to develop magnetic devices with ultra-high density.[4]

One existing challenge for AFM spintronics is the effective regulation of magnetization.[5] Although the AFM materials are essentially magnetic, they do not have macroscopic magnetization due to the alternatively antiparallel magnetic spins.[6] Therefore, AFM materials are very insensitive to the magnetic perturbation.[7] As an alternative, voltage control of magnetism provides a promising way in realizing next generation of fast, compact, and energy-efficient magnetic memories and sensors.[8-10] There are some studies related to the piezoelectric strain-controlled EB with in-plane anisotropy,[9, 11] demonstrating the feasibility of using electric field (E-field) to manipulate AFM moments and domains. E-field induced interfacial strain mediation can also regulate the spin structure of AFM materials[12] with an excellent fatigue durability,[13] which is promising in antiferromagnetic piezo-spintronics. Nevertheless, because that the alternative spins of AFM materials are strongly pinned,[14] the existing E-field control of EB and AFM spins are usually confined at a low temperature[1, 15] or require a large H-field assistance.[12] Although the magnetoelectric (ME) switching of perpendicular EB has been studied by changing the interfacial exchange and magnetization at low temperature,[1] the control of perpendicular EB by strain-mediated ME coupling has not been demonstrated yet. Therefore, we considered perpendicular EB on ferroelectric Pb(Mg$_{1/3}$Nb$_{2/3}$)O$_3$-PbTiO$_3$ (PMN-PT)[16] substrate to realize the electrical modulation of antiferromagnetism.

In this work, we utilized vibrating sample measurements (VSM) and ferromagnetic resonance (FMR) to study the strain-mediated voltage control effect. Both the film fabrication and characterization were carried at room temperature, which has overcome the requirement of a high-temperature magnetic annealing during the EB formation[17] as well as the difficulty of low-temperature testing.[1, 15] For each method, maximal E-field induced EB changes (ΔH$_{eb}$) of 47 Oe and 95 Oe were obtained in the typical perpendicular EB system with the structure of Pt/IrMn /(Co/Pt)$_2$/Ta/PMN-PT. In the thinned Pt/IrMn/Co/Pt/Ta/PMN-PT structure, we report both strain-mediated ME coupling and ionic liquid gating methods for the voltage regulation of perpendicular EB at room temperature. As a result, the charge accumulation in the gating process aroused by the applied voltage generates ~560 Oe hysteresis loop offset, which has not been observed at room temperature. This work exhibits potential applications in energy-efficient AFM devices and high-density memories.

The perpendicular EB structure studied here is Pt 4 nm/IrMn 4 nm/(Co 7 Å /Pt 9.5 Å)$_2$/Ta 3 nm/PMN-PT(011) multiferroic heterostructure, which is a typical perpendicular EB system. The PMA is induced by the (Co/Pt)$_n$ lamination layer while EB is generated at the IrMn/(Co/Pt)$_n$ interface. We used magnet to generate a perpendicular H-field (~1150 Oe) and induced a uniaxial anisotropy at the AFM/FM interface. Based on this structure, we utilized VSM (a popular and direct method for magnetism understanding) and FMR (more accurate for anisotropy quantitation) to systematically study the strain-mediated voltage control effect.

Typical voltage-tuned hysteresis loops of configuration I and II are presented in **Figure 1**. For the (011) cut PMN-PT, the E-field along thickness direction could generate an in-plane compressive strain along the [100] direction ($d_{31}$ = -1800 pC N$^{-1}$, θ= 0° in configuration I) and a tensile strain along the [0-11] direction ($d_{32}$ = 900 pC N$^{-1}$, θ= 0° in configuration II).[18] The charge effect is only active at the FM/ferroelectric (FE) interface[19] and is ineffective for multilayer systems. Besides, charge effect[20] can be excluded in this study because that the 3 nm Ta buffer layer and 0.95 nm Pt layer have separated the bottom Co layer from PMN-PT substrate, preventing interfacial charge accumulation at the Co side. The $H_{eb}^{VSM}$ refers to the center shift of hysteresis loop and Δ$H_{eb}$ refers to the E-field induced change of $H_{eb}^{VSM}$. The positive exchange bias field $H_{eb}$ is defined as a shift toward negative field: $H_{eb} = -(H_{C,left} + H_{C,right})/2$. As shown in Figure 1, the interlayer interaction and the magnetic anisotropy can be changed by the voltage through the strain-mediated ME coupling. At θ= 0° (Figure 1(a)), a significant enhancement of Δ$H_{eb}$ = 47 Oe is observed for configuration I while there is no obvious change of magnetic anisotropy since the squareness of magnetic hysteresis loop is almost the same. As H-field rotates away from the in-plane [100] direction, the magnetic anisotropy is more easily affected by the applied voltage while the exchange bias becomes less sensitive, as shown in Figure 1 (b) and (c). This is probably because that Co 3d - Pt 5d interfacial hybridization in (Co/Pt)$_2$ lamination layer (leads to PMA) and the Co - IrMn interface (leads to in-plane EB) both response to the external voltage.[9, 21, 22] Under an applied voltage, the structural transitions in the ferroelectric PMN-PT substrate produce changes in the stress experienced by the adjacent ferromagnetic Co/Pt multilayer,[23-25] and modified the noncollinear relative orientation at the top Co/IrMn interface.[24] Both the Co 3d - Pt 5d interface[21, 22] and the exchange bias interface[9] are strongly angular dependent. Along the easy axis ($\theta$= 90°), the Co 3d - Pt 5d is more sensitive to voltage tuning than that at 0°, and the modulation of voltage is mainly reflected by the variation of magnetic anisotropy (i.e., the squareness of hysteresis loop). At $\theta$= 0° (hard axis), the voltage effect is mainly reflected by the $\Delta H_{eb}$.

As for configuration II, a similar trend of magnetic hysteresis loops and exchange bias shifts are observed under external E-field. Since the anisotropic piezoelectric coefficients of [01-1] direction is smaller than the [100] direction, the modulation effect along [01-1] is slightly weaker than that along the [100] direction. For both configurations, the magnetic state is stable and would not drift if the applied voltage remains the same.

Our previous work has demonstrated that there are both linear and non-linear piezo response in PMN-PT(011) single crystals.[22, 26] The commonly used linear strain effect is proportional to the applied voltage,[27] while the non-linear strain effect is arising from the 71° and 109° ferroelastic switching which can be generated by a small E-field near the coercive field.[26] We have measured butterfly-like stain curve of the PMN-PT (011) substrate and found that the E-field induced compressive stain (maximum strain

~ -0.5%) is larger than the tensile strain (maximum strain ~ +0.3%).[28] This means that the strain effect of the [100] direction can be more remarkable than that of the [01-1] direction. There are two different mechanisms acting in concert for the implementation of electric control of EB.[24] The first one is that the electrically controlled stress mainly gives rise to changes in the adjacent FM layer, and the accompanying magnetoelastic anisotropy contribution further changes the magnetic anisotropy.[24, 29] The second is that the relative orientation of the FM/AFM interface and the effective exchange at the interface are also controllably affected by the strain.[9, 24] The structural transitions in the (011)-oriented PMN-PT produce changes in the stress experienced by the adjacent ferromagnetic Co/Pt multilayer,[24, 25] therefore, the exchange coupling at the top Co/IrMn interface is strongly modified.

The angular dependences of VSM-measured EB ($H_{eb\text{-VSM}}$) for both configurations are summarized in **Figure 2**. With an applied 10 kV/cm E-field, the shift of EB field $\Delta H_{eb}$ is significant at certain degrees (0°, 15° for I; 15°, 30°, 45° for II) while negligible at other degrees (30°~90° for I). Earlier work by Liu et al.[9] also observed an angular dependence of EB; in their FeMn/$Ni_{80}Fe_{20}$/FeGaB/PZN-PT multiferroic structure, the maximum $\Delta H_{eb}$ of 42 Oe was obtained at 55° along [100] direction and 48 Oe at 60° for [01-1] direction, respectively. Our work demonstrate that the strain-mediated ME coupling not only influences EB, but also the PMA. This exchange-bias field variation under E-field is related to the long-range nature of strain and can be interpreted by the interfacial competition between FM and AFM.[24]

Besides the hysteresis loop measurements, angular dependences of voltage-tuned magnetic properties were studied by FMR at room temperature. **Figure 3** dispalys the results measured by FMR, which provides another characterization method for the EB test.[30-32] Through the resonance fields ($H_r$) at different angles (Figure 3 (a), (c)), EB can be calculated ($H_{eb-FMR}$) according to the following equation, [30, 33]

$$H_{eb}^{FMR}(\varphi_H) = -\frac{1}{2}[H_r(\varphi_H) - H_r(\pi + \varphi_H)] \qquad (1)$$

Using the above expression, we obtain $H_{eb-FMR}$ as shown in Figure 3 (b) and (d). Similar to the $H_{eb-VSM}$, $H_{eb-FMR}$ also has the perpendicular uniaxial symmetry. $H_r$ at 90° is smaller than the $H_r$ at 0°, indicating that the system has PMA. Besides, the obtained $H_{eb-FMR}$ is also strong at certain angles while slight at some other angles, which is in accordance to the voltage modulation process measured by VSM. In the rotational experiments, we obtain maximal $\Delta H_{eb}$ of 95 Oe at θ=10° along the [100] direction. One can also notice that the amplitude and critical angle measured by the two methods are different. As shown in **Figure 4**, the unidirectional exchange anisotropy values measured by out-of-plane FMR are larger than the loop shifts measured via VSM. To eliminate the measurement error, each hysteresis loop/spectrum was measured for three times. Besides results at 10 kV/cm, the results at lower voltages display transitional intermediate states, indicating that the changes of exchange bias under applied voltages truly reflect inner physical variations. Although both $H_{eb-VSM}$ and $H_{eb-FMR}$ are caused by the uniaxial symmetry, they are reflected by different magnetization processes, which can cause different $H_{eb}$ values.[30-32] VSM method needs to reverse the magnetization; the encounter of energy barriers make the AFM domain structure relaxes to a new energy minimum state during the descending branch, which will not recover during the ascending branch.[34] This has been demonstrate as a rotatable anisotropy.[34] Contrarily, the FMR is a perturbative method with only a small amount of magnetization is involved, which will not form a new energy state.[31, 34] Besides, FMR method is dependent on microwave frequency, which means different modes provide different values.[30] Therefore, it is predictable that the measured exchange anisotropy of the two method have different values.

For an in-plane EB system, the $H_{eb}$ value measured through hysteresis is about 20% larger than that of the FMR method,[34] which is contrary to the tendency observed here (the $H_{eb-FMR}$ is 0.5~2 times larger than the $H_{eb-VSM}$), suggesting different anisotropy distribution for in-plane and out-of-plane EB systems. The strain-mediated voltage control effect is a combination of EB and PMA; the perpendicular EB regulation in Pt/IrMn/(Co/Pt)$_2$/Ta nanostructure has a limitation. For the voltage regulation of interface phenomena,[35] our previous studies have demonstrated that when the film is thin enough, the control effect can be rather effective.[8, 36] To let the voltage induced lattice distortion transmit effectively to the IrMn/Co interface instead of to the (Co/Pt)$_n$ interlayers, we further thinned the EB system with minimum Co/Pt lamination layer.

Therefore, based on the thinned EB system with the structure of Pt/IrMn (t nm)/Co (1.2 nm)/Pt (2 nm)/Ta (3 nm), we tried both strain-mediated ME coupling and ionic liquid gating methods. We chose (001)-oriented PMN-PT to be the ferroelectric substrate, which is suitable and simple for the study of voltage controllable magnetism at out of plane.[21] To simplify the situation, we utilized VSM to study the magnetic properties at the perpendicular direction. Basic magnetic properties of the new multiferroic EB system were shown in **Figure 5**. The magnetic hysteresis loops can be regulated by the thickness of IrMn layer (Figure 5a); the intensity of EB increases with the IrMn thickness until reaches the saturation point, in which the AFM grain volume is large enough to pin all the FM spins at the interface.[37] The results shown in figure 5 (b) demonstrate that the decrease in Pt thickness is acceptable since that the variation trend of IrMn thickness is consistent with figure 5 (a). Therefore, we further reduce the thickness of cap Pt layer to minimize the film thickness and enhance the interfacial gating effect. For the thinned EB system, domain nucleation occurs at the switching field, also showing an asymmetric flipping behavior (Figure 5 (c-h)).

The results of electrical modulation of EB are shown in **Figure 6**. For the strain effect, we studied two samples with $t_{IrMn}$= 6 nm and 4.5 nm. At the perpendicular direction, the hysteresis loops exhibit maximum perpendicular EB shifts of ~ 35 Oe under small opposite E-fields (34 Oe for 6 nm, 36 Oe for 4.5 nm); while $\Delta H_{eb}$ under 8 kV/cm E-field is ~18 Oe (21 Oe for 6 nm, 16 Oe for 4.5 nm). The hysteresis loops have offsets without obvious changes in squareness, indicating that the strain effect transmits effectively to the IrMn/Co interface. The small E-field near the coercive field (1.5 kV/cm)[26] can give rise to the 71° and 109° ferroelastic switching, in which the non-linear piezo response generates a larger out-of-plane lattice strain than the linear strain effect. In other words, the non-linear strain effect is more efficient than the commonly used linear strain effect where the strain is in direct proportion to the external voltage.[26, 27] This phenomenon is not obvious in the Pt/IrMn/(Co/Pt)$_2$/Ta/PMN-PT (011) multiferroic heterostructure that we discussed earlier, in which the piezo response at high E-field (10 kV/cm) is more clear to identify.

The gating method illustrated in Figure 6 (d) realizes ME coupling via E-field induced charge accumulation (details can be found in the methods section).[8, 36, 38] Under an external E-field, anions and cations in the ionic liquid (IL) accumulate at the top and bottom surfaces, respectively. At the IL/film interface, there are opposite charge accumulations forming an electric double layer (EDL) with a thickness ~ 3 nm.[13, 28] As demonstrated by our previous work, the ultra-high surface charge density (up to $10^{15}$/cm) in EDL could lead to fascinating interfacial ME phenomena.[8, 36, 39] Hysteresis loops shown in Figure 6 (e) display a large offset and the shift in Figure 6 (f) can be up to 560 Oe when the IrMn thickness decreases from 6 nm to 5 nm. The gating method provides a room-temperature technique to enhance voltage control effect of the perpendicular EB, which is even more remarkable than the previous results[1] of ionic liquid gating confined at low temperature. Nevertheless, the gating method suffers from a recovery problem as well as a slow response rate, which should be improved in future study.

In summary, Pt/IrMn/(Co/Pt)$_2$/Ta/PMN-PT multiferroic nanostructure with perpendicular EB was fabricated without annealing process, and the voltage regulation of perpendicular EB was studied at room temperature. Maximal E-field induced $\Delta H_{eb}$ of 47 Oe and 95 Oe were obtained along the [100] direction by using VSM and FMR methods, respectively. Due to the different measurement principle, the obtained $H_{eb\text{-}FMR}$ was 0.5~2 times larger than the $H_{eb\text{-}VSM}$. A thinner EB system with Pt/IrMn/Co/Pt/Ta structure was also investigated on both PMN-PT (001) and Si/SiO$_2$ substrates, and the corresponding voltage-induced non-linear strain effect and charge accumulation can generate $\Delta H_{eb}$ in perpendicular direction without obvious change of anisotropy, respectively. The understanding of the room-temperature voltage regulation process for the perpendicular EB system is of great technological significance for the applications in energy-efficient and high-density AFM spintronic devices.

**Experimental Section**

*Sample growth:* Pt 4 nm/ IrMn 4 nm/(Co 7 Å /Pt 9.5 Å)$_2$/Ta 3 nm/PMN-PT (011) multiferroic heterostructure and Pt/IrMn (t nm)/Co (1.2 nm)/Pt (2 nm)/Ta (3 nm) heterostructures on PMN-PT (001) and Si/SiO$_2$ substrates were deposited by magnetron sputtering at room temperature. With a perpendicular H-field (~1150 Oe) across the substrate thickness direction, providing by the permanent magnet, we managed to induce exchange coupling between Co/Pt and IrMn layer without annealing process. The base pressure was less than 10$^{-7}$ Torr; the working Ar pressure was 3 mTorr; DC power was 20-30 W. PMN-PT substrates were used without pre-poling, and we tested the initial virgin states and the poling states with applied E-field for comparison.

*Device Fabrication:* For the strain-mediated ME coupling multiferroic heterostructures. The voltages were applied by the 6517B Electrometer. Copper wires were used as the connecting line between surface film plane (i.e. top Pt layer) and 6517B Electrometer. Ta (100 nm) and Pt (100 nm) layers were sputtered on the bottom of PMN-PT as an electrode, and the in situ E-fields were applied through the thickness of PMN-PT substrate. Device details can be found in Figure 1 (configuration I and II).

For the ionic liquid gating regulation, its configuration is shown in Figure 6 (d).[8, 36, 38] The ionic liquid [DEME]$^+$[TFSI]$^-$ was dropped onto the film plane directly contacting the sample and the gold wire. The gold wire was used to avoid electrochemical reaction. The gold electrode and film electrode were isolated to each other to ensure that the voltage was applied across the ionic liquid. Gating voltage was applied by the B2901A Electrometer. Before formal measurements, we waited for at least 5 minutes to ensure charge balance.

*Magnetic property measurements:* In situ voltage manipulation was carried out by using VSM (Lake Shore 7404), FMR (JEOL, JES-FA200) and MOKE microscope (Evico Magnetics, em-Kerr-Highres). The external voltage was applied by 6517B electrometer. All the characterizations were performed at room temperature.

**Acknowledgment:**
The work was supported by the National Key R&D Program of China (Grant No. 2018YFB0407601), the Foundation of National Key Laboratory (Grant No. 2018SSFNKLSMT-04), the Natural Science Foundation of China (Grant Nos. 51472199, 11534015, 51602244, and 51802248), the Key R&D Program of Shaanxi Province (Grant No. 2018GY-109), the National 111 Project of China (B14040), and the Fundamental Research Funds for the Central Universities (xjj2018207).

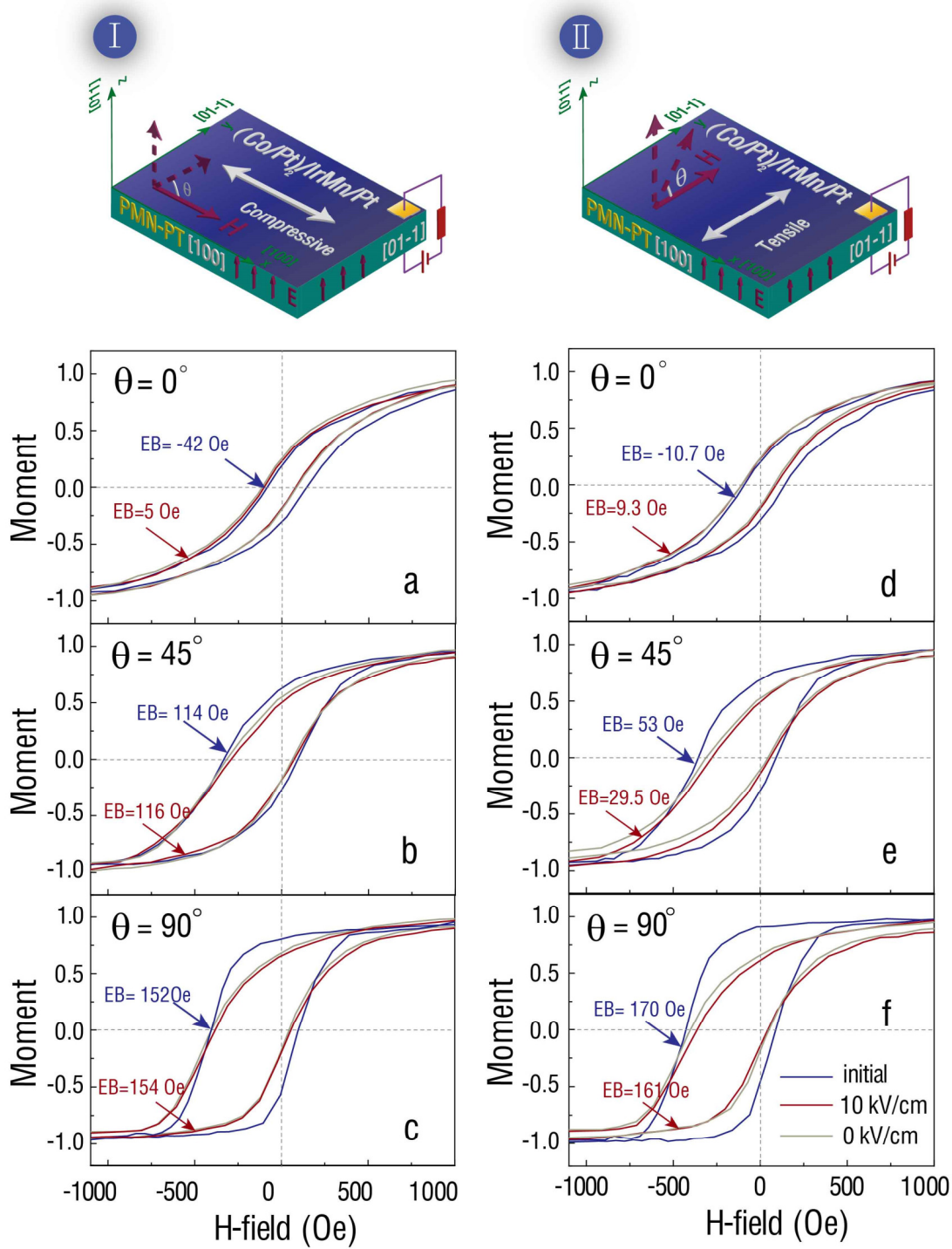

**Fig. 1** Angular dependence for Pt (4 nm)/IrMn (4 nm)/(Co 0.7 nm/Pt 0.95 nm)$_2$/Ta (3 nm)/PMN-

PT (011) multiferroic heterostructure. (a–c) Magnetic hysteresis loops with and without external voltage at θ =0°, 45°, 90° in configuration I. (d–f) Magnetic hysteresis loops at θ=0°, 45°, 90°for configuration II. θ is the angle between external magnetic field H and the film plane, x and y represent the [100] and [01-1] crystallographic direction of PMN-PT, respectively. For configuration I, H-filed is rotate around the [100] direction; for configuration II, H-filed is rotate around the [01-1] direction. In both configurations, θ=0° represents the in-plane direction, while θ=90° refers to the out-of-plane direction. The E-field is applied across the thickness direction; since the thickness of PMN-PT substrate is 0.5 mm, the applied 500 V corresponds to a 10 kV/cm E-field intensity. The PMN-PT substrates were used without pre-poling and we compared the initial virgin states and the poling states.

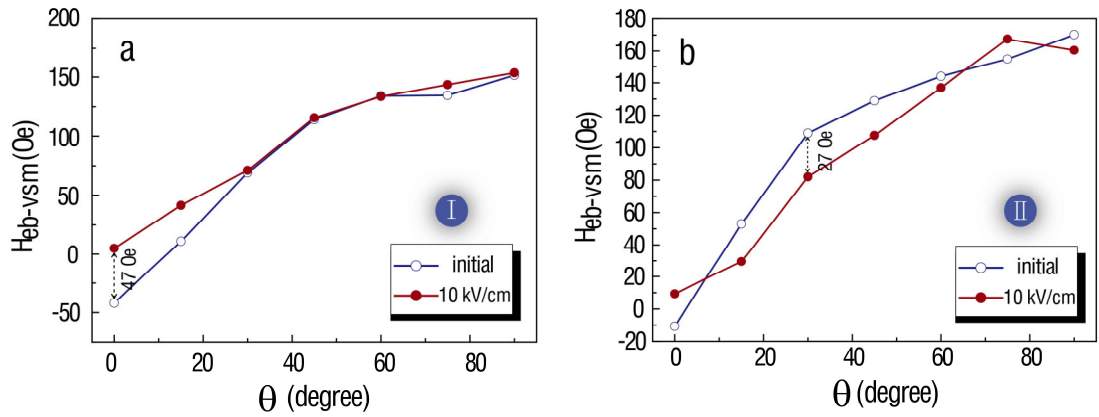

**Fig. 2** Angular dependence of exchange bias measured with VSM while under various E-fields, based on configuration I (a) and configuration II (b), respectively.

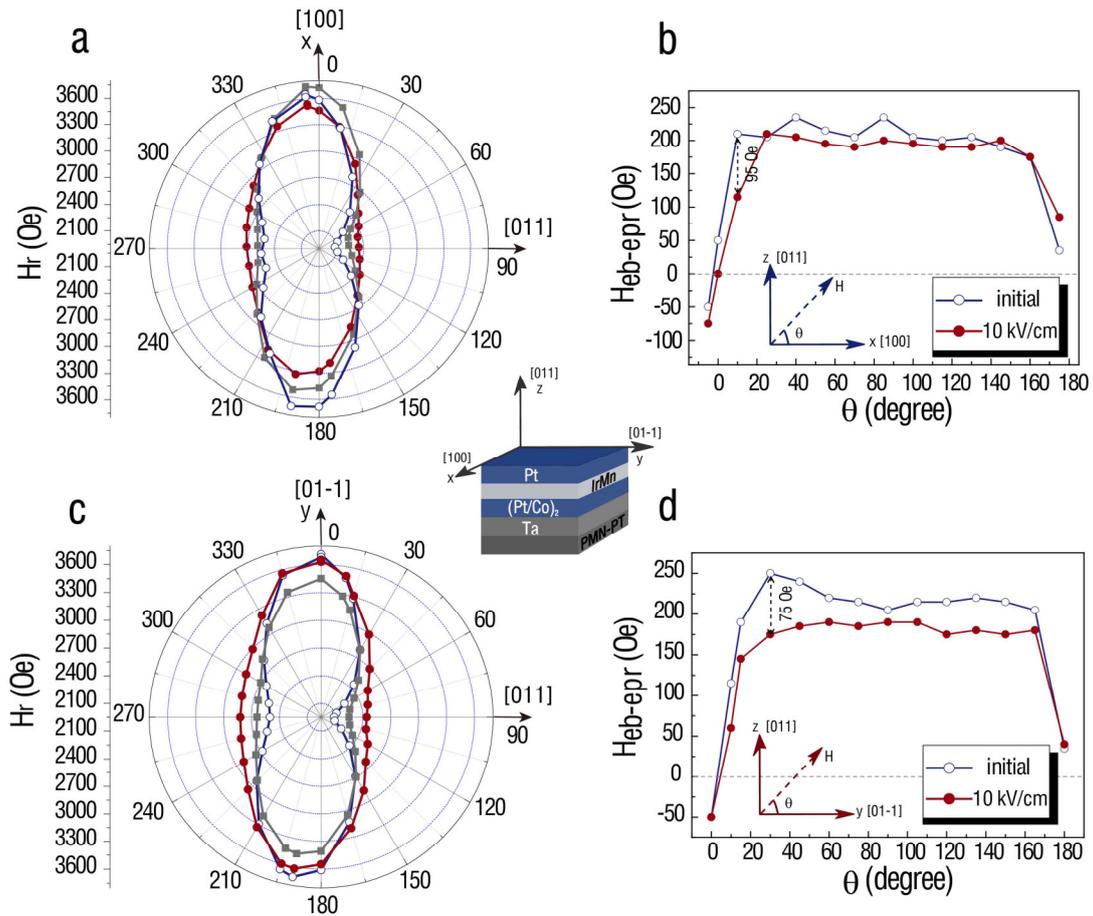

**Fig. 3** Angular dependence of FMR (a, c) and exchange bias (b, d) for Pt 4 nm/ IrMn 4 nm/(Co 7 Å /Pt 9.5 Å)$_2$/Ta 3 nm/PMN-PT (011) multiferroic heterostructure based on configuration I (a, b) and configuration II (c, b) respectively.

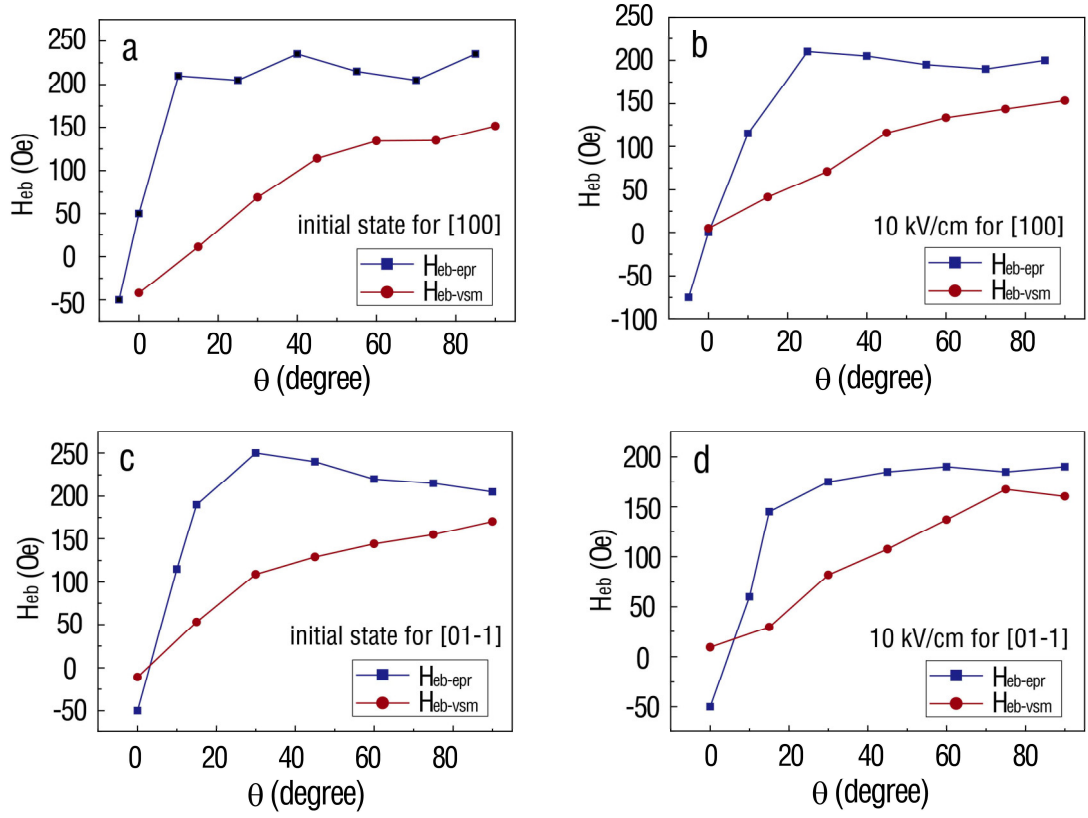

**Fig. 4** (a) Comparison of $H_{eb}^{epr}$ (black) and $H_{eb}^{vsm}$ (red) at initial state (a, c) and E=10 kV/cm (b, d) for Pt 4 nm/IrMn 4 nm/(Pt 9.5 Å/Co 7 Å)$_2$/Ta 3 nm/PMN-PT (011) multiferroic heterostructure.

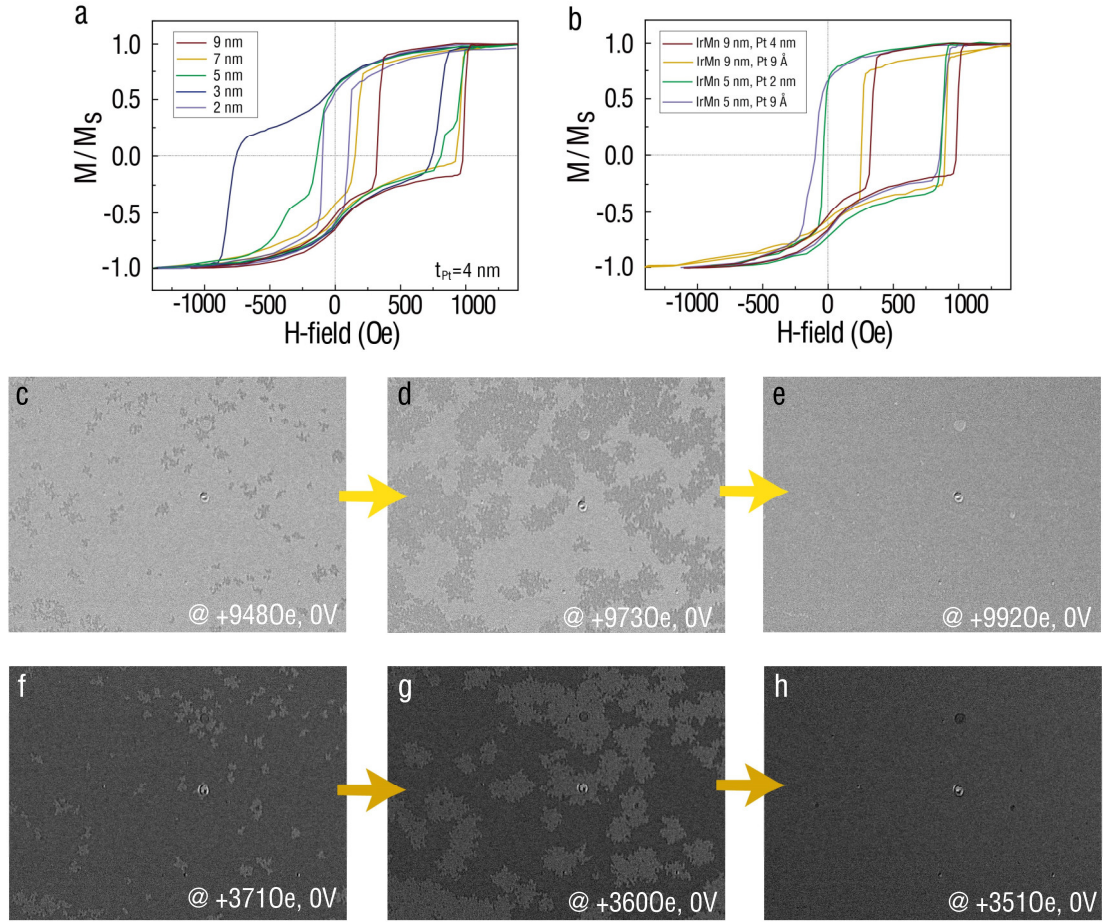

**Fig. 5** (a) are magnetic hysteresis loops for Pt (4 nm)/IrMn (t nm)/Co (1.2 nm)/Pt (2 nm)/Ta (3 nm)/PMN-PT (001) multiferroic heterostructures with different IrMn thickness. (b) are hysteresis loops with varied IrMn and top Pt thickness. (c-e) Domain nucleations when H-field increases from +948 Oe to +992 Oe. Figure (f-h) are the MOKE images of domain switching between +371 Oe and +351 Oe.

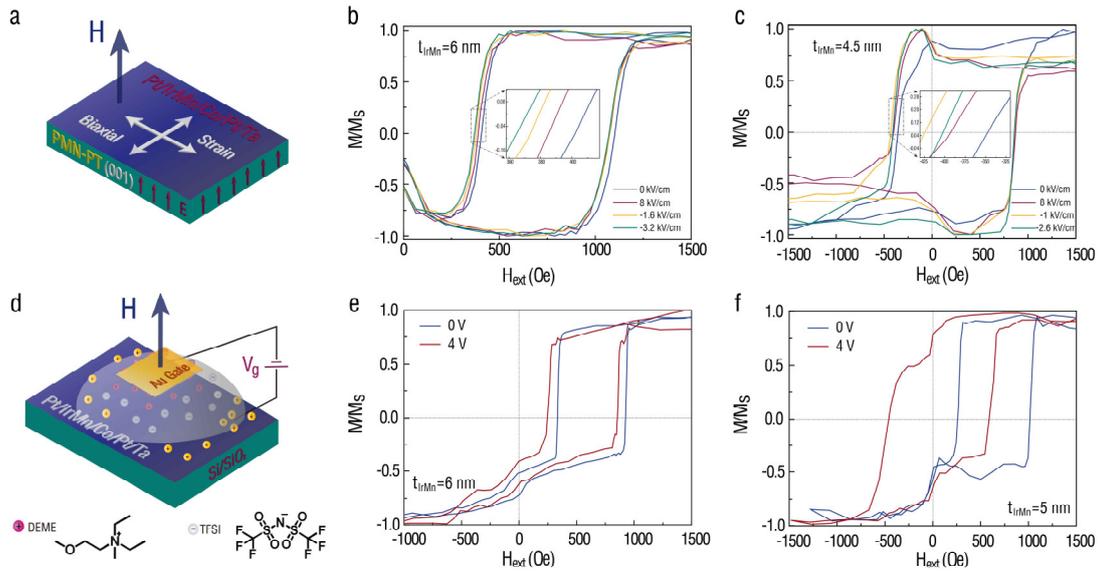

**Fig. 6** Voltage regulation for the thinned perpendicular EB systems. (a) Schematic for the strain-mediated ME coupling in Pt (9 Å)/IrMn (t nm)/Co (1.2 nm)/Pt (2 nm)/Ta (3 nm)/PMN-PT (001) multiferroic heterostructures. (b, c) Corresponding magnetic hysteresis loops at $t_{IrMn}$= 6, and 4.5 nm for the strain regulation. The enlarged inserted pictures show the relative shifts of hysteresis loops under different E-fields. (d) Schematic for the ionic liquid gating in Pt (9 Å)/IrMn (t nm)/Co (1.2 nm)/Pt (2 nm)/Ta (3 nm)/Si/SiO$_2$ heterostructures. (e, f) Corresponding magnetic hysteresis loops at $t_{IrMn}$= 6, and 5 nm for ionic liquid gating.